\documentclass{emulateapj}

\begin{document}

\title{Planet formation around stars of various masses: The snow line
  and the frequency of giant planets}

\shorttitle{SNOW LINE AND PLANET FREQUENCY}
\shortauthors{KENNEDY \& KENYON}

\author{Grant M. Kennedy\altaffilmark{1}}
\affil{Research School of Astronomy and Astrophysics, Mt Stromlo
  Observatory, Australian National University, ACT 2611, Australia}
\affil{ANU Planetary Science Institute, Australian National
  University, Canberra, Australia}
\email{grant@mso.anu.edu.au}

\and

\author{Scott J. Kenyon}
\affil{Smithsonian Astrophysical Observatory, Cambridge, MA 02138, USA}
\email{kenyon@cfa.harvard.edu}

\altaffiltext{1}{\emph{current address}: Smithsonian Astrophysical
  Observatory, Mail Stop 16, 60 Garden St, Cambridge, MA 02138, USA}
  
\begin{abstract}
  We use a semi-analytic circumstellar disk model that considers
  movement of the snow line through evolution of accretion and the
  central star to investigate how gas giant frequency changes with
  stellar mass. The snow line distance changes weakly with stellar
  mass; thus giant planets form over a wide range of spectral
  types. The probability that a given star has at least one gas giant
  increases linearly with stellar mass from 0.4\,$M_\odot$ to
  3\,$M_\odot$. Stars more massive than 3\,$M_\odot$ evolve quickly to
  the main-sequence, which pushes the snow line to 10--15\,AU before
  protoplanets form and limits the range of disk masses that form
  giant planet cores. If the frequency of gas giants around solar-mass
  stars is 6\%, we predict occurrence rates of 1\% for
  0.4\,$M_\odot$ stars and 10\% for 1.5\,$M_\odot$ stars. This
  result is largely insensitive to our assumed model
  parameters. Finally, the movement of the snow line as stars
  $\gtrsim$2.5\,$M_\odot$ move to the main-sequence may allow the
  ocean planets suggested by \citeauthor{2004Icar..169..499L} to form
  without migration.
\end{abstract}

\keywords{planetary systems: formation --- planetary systems:
  protoplanetary disks --- stars: evolution --- stars: formation}

\section{Introduction}\label{sec:intro}

In the last ten years, the discovery of more than 200 extra-solar
planets,\footnote{http://vo.obspm.fr/exoplanetes/encyclo/encycl.html}
and more than 200 debris
disks,\footnote{http://www.roe.ac.uk/ukatc/research/topics/dust/identification.html
} suggests that planet formation is a common and robust
process. Planet masses inferred from debris disks range from
terrestrial to Jovian, at distances as great as tens of AU from the
central star
\citep[e.g.][]{2004AJ....127..513K,2005ApJ...619L.187G}. The nature
and sensitivity of radial velocity surveys means that most of the
planets are $\sim$Jupiter mass gas giants in close orbits around
Sun-like stars. However, recent discoveries as diverse as icy
$\sim$Neptune-mass planets orbiting M dwarfs
\citep[e.g.][]{2005ApJ...634..625R}, and debris disks around A-type
stars \citep[e.g.][]{2005ApJ...620.1010R} show that planet formation
occurs over a wide range of spectral types.

Current theory suggests that planets form in similar ways around all
stars. Thus, the increasing diversity of stellar hosts and planetary
systems provides an opportunity to test these theories. For this
reason, the types of planets most likely to form around stars of
differing spectral types has become a renewed area of study
\citep[e.g.][]{2005ApJ...626.1045I,2006ApJ...643..501B,2006A&A...458..661K,2006ApJ...650L.139K},
after the idea was first explored by \citeauthor{1988MNRAS.235..193N}
nearly 20 years ago
\citep{1987MNRAS.224..107N,1988MNRAS.230..551N,1988MNRAS.235..193N}.

Theories of Solar System formation generally include the ``snow
line,'' where ices condense from the nebular gas. The snow line
distance is usually fixed in a disk with a time independent surface
density and temperature profile around a main-sequence star
\citep[e.g.][]{2005ApJ...626.1045I}. In a more realistic picture, the
disk and stellar properties evolve considerably during the 1--10\,Myr
pre--main-sequence (PMS) lifetime when planets probably form
\citep[e.g.][]{1987Icar...69..249L,1996Icar..124...62P}. As the disk
temperature evolves with time, movement of the snow line may therefore
influence the properties of theoretical planetary systems
\citep[e.g.][]{2006ApJ...650L.139K,2007ApJ...654..606G}.

Here, we begin to develop a time dependent model for the formation of
gas giant cores that considers the PMS evolution of the star and
surrounding accretion disk. We introduce a simple semi-analytic disk
model, based on the ``minimum mass solar nebula,'' that links movement
of the snow line through evolution of disk accretion and stellar
luminosity. In contrast to previous studies
\citep[e.g.][]{2005ApJ...626.1045I,2006A&A...458..661K}, our analysis
suggests that gas giant formation around stars more massive than the
sun is more likely than around less massive stars.

We cover the background important to our story in \S
\ref{sec:background}, consider the snow line in \S
\ref{sec:motivation}, and outline our model in \S \ref{sec:model}. We
present our results in \S \ref{sec:results}, and discuss and conclude
in \S \ref{sec:discussion} and \S \ref{sec:summary}.

\section{Background}\label{sec:background}

Planetary systems form in circumstellar disks, which evolve on
timescales comparable to the pre--main-sequence (PMS) stage of stellar
evolution. Observations indicate a wide range of disk masses $M_{disk}
\sim 0.01$---0.1\,$M_\star$ \citep[where $M_\star$ is the stellar
mass,
e.g.][]{1995ApJ...439..288O,2000prpl.conf..559N,2005ApJ...631.1134A,2006ApJ...641.1162E,2006ApJ...645.1498S}
and radii $\sim$100--1000\,AU \citep{1996AJ....111.1977M}. The
lifetime of the primordial, optically thick, dusty component of the
disk is $\lesssim$10\,Myr, with a median timescale of $\sim$3\,Myr
\citep[e.g.][]{1993prpl.conf..837S,2001ApJ...553L.153H}. Though harder
to observe, the gaseous component of the disk is probably removed by
viscous accretion \citep{1974MNRAS.168..603L} and photoevaporation
\citep[e.g.][]{2000prpl.conf..401H,2004ApJ...611..360A,2006MNRAS.369..229A}
on similar timescales \citep{1995Natur.373..494Z,2006ApJ...651.1177P}.

These timescales place strict observational limits on the important
stages of planet formation. Planetesimals must form rapidly to enable
further grain growth and protoplanet formation by coagulation
\citep[e.g.][]{1969QB981.S26......}. To attract significant
atmospheres and form gas giants, protoplanets need to reach
masses of 5-10\,$M_\oplus$
\citep[e.g.][]{1996Icar..124...62P,2000ApJ...537.1013I} before the
nebular gas is removed.

In coagulation models, dust particles on near circular orbits with
small relative velocities grow through repeated collisions and mergers
in circumstellar disks. Further dynamical evolution through
``runaway'' \citep{1989Icar...77..330W,1996Icar..123..180K} and
``oligarchic'' \citep{1998Icar..131..171K} growth leads to
``isolated'' protoplanets, whose mass $M_{iso}$ and spacing depend on
their radial distance $a$ from the central star via the Hill radius
$R_H = a \left( M_{iso} / 3 M_\star \right)^{1/3}$
\citep[e.g.][]{1987Icar...69..249L,2007prpl.conf..591L}
\begin{equation}\label{eq:miso}
  M_{iso} = \frac{ \left( 4 \pi B \sigma a^2 \right)^{3/2} }
  { (3 M_\star)^{1/2} } \, ,
\end{equation}
where $\sigma$ is the disk surface density. Protoplanets are spaced at
$2 B R_H \sim 8 R_H$ intervals \citep{1998Icar..131..171K}.  Used in
combination with equation (\ref{eq:miso}), the ``minimum mass solar
nebula'' \citep[MMSN,][]{1977Ap&SS..51..153W,1981PThPS..70...35H} with
$\sigma \propto a^{-\delta}$ (where $\delta = 1$--1.5), gives a simple
model of protoplanet formation.

The ``snow line''---the point in the disk that separates the inner
region of rocky planet formation from the outer region of icy planet
formation---is an important feature of the MMSN
\citep[e.g.][]{2000ApJ...528..995S,2005ApJ...626.1045I,2006Icar..181..178C}. Condensation
of ices outside the snow line increases the disk surface density by a
factor $f_{ice} \sim 3$,\footnote{The usual value is $\sim$4, but
  recent solar abundance figures for oxygen
  \citep{2005ASPC..336...25A} indicate 3 is more reasonable. Recent
  composition data from 9P/Tempel 1 may argue for an even lower
  ice/rock ratio \citep{2005Natur.437..987K}.} which leads to factor
of 5 larger isolation masses (eq. \ref{eq:miso}). In an MMSN model
with $\sigma = 10$ g\,cm$^{-2}$ at 1\,AU and $\delta=3/2$, $M_{iso}
\approx 0.1$ (1)\,$M_\oplus$ at 1 (5)\,AU. To achieve the probable
core mass of 5--10\,$M_\oplus$ for Jupiter \citep{2004ApJ...609.1170S}
the MMSN can be augmented beyond the snow line by a factor
$\sim$4. Alternatively, if $\delta = 1$ then $M_{iso} \approx
5\,M_\oplus$ at 5\,AU. Models that relax the assumption of a smooth
radial profile find an enhanced surface density near the snow line
\citep[e.g.][]{2004ApJ...614..490C,2006Icar..181..178C}. A common
theme among both MMSN and more detailed models is the surface density
added by ice condensation.

Because the timescale for planet growth is $t \propto P / \sigma
\propto a^3$ for $\sigma \propto a^{-3/2}$, where $P$ is the orbital
period \citep[e.g.][see also
\citet{2004ApJ...614..497G}]{1987Icar...69..249L}, ice condensation
also leads to shorter growth times. Numerical simulations by
\citet{2004AJ....127..513K,2004ApJ...602L.133K} find the time to form
1000--3000\,km objects agrees with this relation. Numerical estimates
of the time to form the Jovian core range from
$\sim$10$^5$--10$^6$\,yr
\citep[e.g.][]{1987Icar...69..249L,1996Icar..124...62P,2003Icar..166...46I,2006ApJ...652L.133C}. In
general, the time to reach isolation $t_{iso}$ provides an estimate of
whether protoplanets form early enough to accrete gas and become giant
planets. Short gas disk lifetimes
\citep[e.g.][]{1995Natur.373..494Z,2006ApJ...651.1177P}, imply a
relatively short isolation time, and place strong constraints on the
time to form gas giants by core accretion.

Gas giant formation by core accretion occurs when protoplanet core
masses are sufficient to attract gas from the nebula. The core mass
sets the timescale for gas giant formation
\citep{2000ApJ...537.1013I,2005Icar..179..415H}. Cores with masses
smaller than $\sim$5\,$M_\oplus$ attract atmospheres
\citep[e.g.][]{2003A&A...410..711I}, but are unable to form a gas
giant before the nebular gas is removed on timescales of
1--10\,Myr. Beyond the snow line the critical core mass where
significant gas accretion occurs is $M_{crit} \sim 7\,M_\oplus \,
\dot{M}_{core}^{0.25} \, \kappa^{0.25}$ \citep[where $\dot{M}_{core}$
is the rate at which planetesimals are accreted onto the core in units
of $10^{-7}\,M_\oplus$ yr$^{-1}$, and $\kappa$ is the grain opacity in
units of cm$^2$ g$^{-1}$,][see also
\citet{2006ApJ...648..666R}]{2000ApJ...537.1013I}. The critical core
mass required to form Jupiter in several Myr \citep[$M_{iso} \sim
5$--10\,$M_\oplus$,][]{1996Icar..124...62P,2005Icar..179..415H}---which
implies $\sigma \sim 10$\,g cm$^2$ at 5\,AU---is consistent with the
core mass inferred from current structural models
\citep[e.g.][]{2004ApJ...609.1170S}.

\subsection{Previous Work}

Most planet formation theories are based on a static MMSN disk around
a solar-mass star. There are several motivating factors for extending
these theories to a range of stellar masses: (i) the increasing
stellar mass range of extra-solar planet hosts, (ii) observed trends
with stellar mass, such as accretion rate and disk mass, and (iii)
theoretical relations with variables that change with stellar mass,
such as orbital period and isolation mass. This extension of Solar
System theory to a range of spectral types began with a series of
papers by \citeauthor{1988MNRAS.235..193N} nearly twenty years ago
\citep{1987MNRAS.224..107N,1988MNRAS.230..551N,1988MNRAS.235..193N}.
More recently, \cite{2006A&A...458..661K} considered formation of
planets around stars of various masses \emph{in situ}, while
\cite{2005ApJ...626.1045I} examined observable planetary systems
resulting from type II migration.

\citeauthor{2006A&A...458..661K} consider disk evolution prior to the
growth of large objects. In their models, the increased inward
migration rate for planetesimals around low-mass stars results in
higher absolute surface densities from 0.1--100\,AU at 1\,Myr. Thus
low-mass stars are more likely to form giant planets. This result is
influenced by their choice of an approximately constant initial disk
mass for all stellar masses. They do not consider planet formation
beyond 5\,AU.

\citeauthor{2005ApJ...626.1045I} base their Monte-Carlo study on the
MMSN. Type II migration---where a planet with sufficient mass opens a
gap in the disk and whose orbit is subsequently coupled to the viscous
evolution of the disk \citep[e.g.][]{1985prpl.conf..981L}---is central
to their model. In their attempt to reproduce the observed
distribution of extra-solar planets, they find that close-in icy
Neptune-mass planets should be much more common than close-in
Jupiter-mass planets around M dwarf stars \citep[see
also][]{2004ApJ...612L..73L}. In contrast to the
\citeauthor{2006A&A...458..661K} study, they find that the likelihood
of a star harbouring gas giants increases with stellar mass up to
solar-mass stars. Their results are influenced by scaling the snow
line distance as $a_{snow} \propto M_\star^2$, based on the
main-sequence luminosity $L_\star \propto M_\star^4$. As we show
below, this simplification places the snow line too close to (far
from) the central star for stars with masses less than (greater than)
a Solar mass when protoplanets form.

In this paper we consider movement of the snow line as disk accretion
subsides and the central star evolves to the main-sequence. Using our
prescription for the snow line position over a range of stellar and
disk masses, we locate regions where gas giant cores form. Assuming
stars are born with disks from a distribution of masses, we then
predict how gas giant frequency varies with stellar mass.

\section{Location of the Snow Line}\label{sec:motivation}

In this section we consider evolution of the disk mid-plane
temperature, and the snow line distance, with a simple model that
includes accretion and PMS evolution. In particular, we are interested
in the stellar mass dependence, rather than a detailed derivation for
a single star. As we show in \S \ref{sec:results}, $\sim$1\,AU
differences between our model and more detailed treatments
\citep[e.g.][]{2000ApJ...528..995S,2006ApJ...640.1115L} do not affect
our conclusions.

The disk mid-plane, where the gas density is highest, is probably
where most ices condense, and has a temperature
\begin{equation}\label{eq:tmid}
  T_{mid}^4 = T_{mid,accr}^4 + T_{irr}^4 \, ,
\end{equation}
where $T_{mid,accr}$ is the mid-plane temperature arising from viscous
forces within the disk, and $T_{irr}$ is temperature due to external
irradiation of the disk by the central star.

The effective disk temperature from viscous accretion is
\citep{1974MNRAS.168..603L}
\begin{equation}\label{eq:taccr}
  T_{eff,accr}^4 = \frac{3}{8 \pi} \frac{G M_\star \dot{M}}{\sigma_{sb} a^3}
\left( 1 - \sqrt{ \frac{ R_\star }{ a } } \right) \, ,
\end{equation}
where $\dot{M}$ is the accretion rate, $R_\star$ is the stellar
radius, and $\sigma_{sb}$ is Stefan's constant. In optically thick
regions near the snow line, the mid-plane temperature is
$T_{mid,accr}^4 \sim 3\tau T_{eff,accr}^4/8$
\citep{1990ApJ...351..632H}, where $\tau = \kappa \sigma_g/2$. The
opacity $\kappa$ is a function of temperature
\citep{1994ApJ...427..987B}, and the gas surface density $\sigma_g$ is
100 times greater than that of solids. The figure of 100 is used in
converting mm dust observations to total disk masses
\citep[e.g.][]{2000prpl.conf..559N} based on the interstellar gas/dust
ratio, and is similar to the solar metallicity (Z) fraction of 0.0122
\citep{2005ASPC..336...25A}. The accretion rate varies with stellar
mass as approximately $\dot{M} \propto M_\star$ for the range of
stellar masses we consider
\citep[0.2--4\,$M_\star$,][]{2003ApJ...592..266M}, and with time as
$\dot{M} \propto \left( t / 10^6\,{\rm yr}
\right)^{-\gamma}$. \citet{1998ApJ...495..385H} derive $\gamma =
1.5$--2.8. The uncertainty is due to the limited age range of their
sample and a large range of accretion rates at a given age. The value
$\gamma = 1.5$ is their ``preferred result.'' We scale $\dot{M}$ with
surface density, which accounts for the observed trend with stellar
mass (if disk mass scales linearly with stellar mass, see \S
\ref{sec:model}), and is consistent with expected viscous evolution
(where $\dot{M} \propto \nu \sigma$ and $\nu$ is the disk
viscosity). For $\sim$1\,Myr old Solar-type stars $\dot{M} \sim
10^{-8}\,M_\odot$ yr$^{-1}$ \citep{1998ApJ...495..385H}. We set
$\dot{M} = 10^{-8}\,M_\odot$ yr$^{-1}$ for an initially three-fold
enhanced MMSN disk, as this disk decays to the ``typical'' observed
$\sim$MMSN mass disk by $\sim$1\,Myr \citep{1998ApJ...495..385H}.

A more complete treatment of the optical depth to the mid-plane would
include evolution of the gas surface density, allowing the mid-plane
temperature to drop somewhat faster than described above as $\sigma_g$
decreases in the inner disk \citep{1974MNRAS.168..603L}. The solid
surface density, which largely resides near the mid-plane and
determines protoplanet characteristics, remains largely unaffected by
the gas disk evolution (aside from snow line evolution).

The disk temperature contribution from irradiation is
\begin{equation}\label{eq:tirr}
   T_{irr} = T_\star \left( \frac{ \alpha }{ 2 } \right)^{1/4}
   \left( \frac{ R_\star }{ a } \right)^{3/4} ,
\end{equation}
where $\alpha \approx 0.005/a_{\rm AU} + 0.05 a_{\rm AU}^{2/7}$ for a
flared disk in vertical hydrostatic equilibrium
\citep[e.g.][]{1986ApJ...308..836A,1987ApJ...323..714K,1997ApJ...490..368C}.
Here $a_{\rm AU}$ is $a$ in units of AU. When the disk is optically
thick to radiation at this temperature, $T_{irr}$ is approximately the
interior temperature for a flared disk \citep{1997ApJ...490..368C}.

\begin{figure*}
  \plottwo{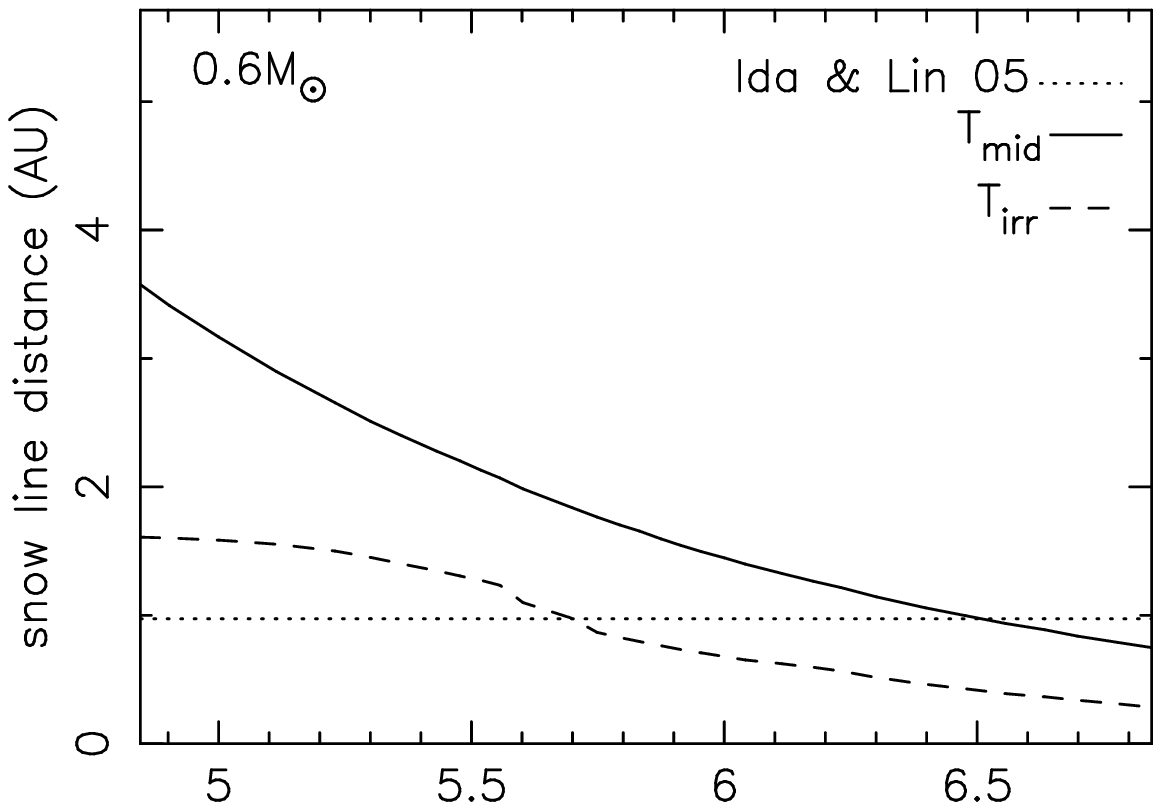}{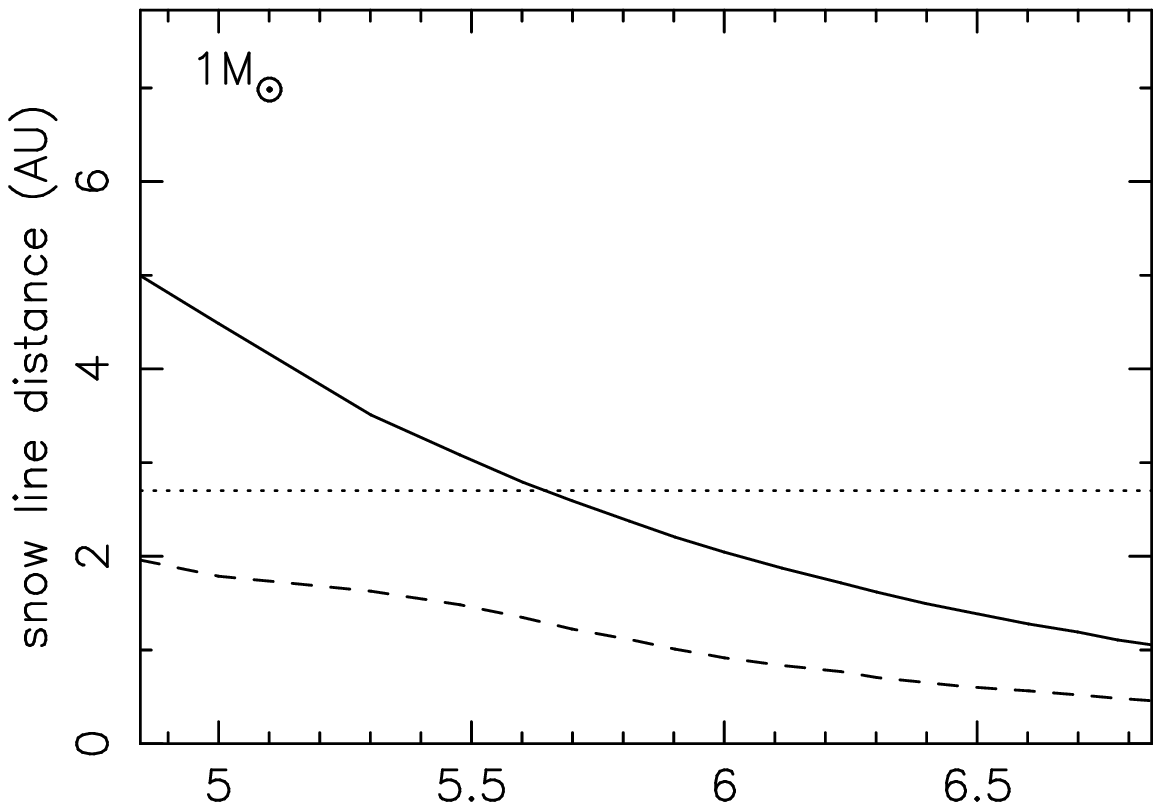}
  \vspace{10pt}\\
  \plottwo{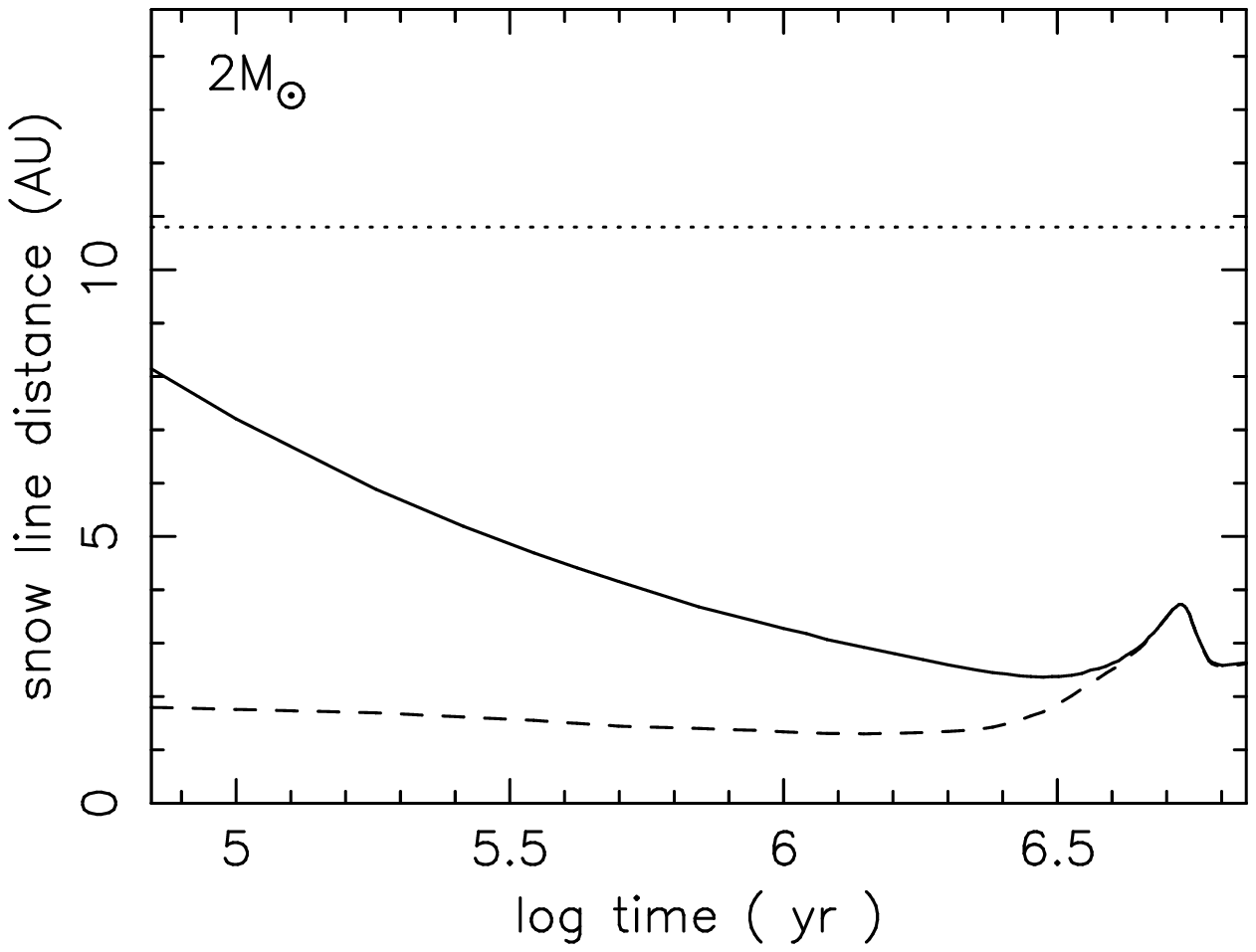}{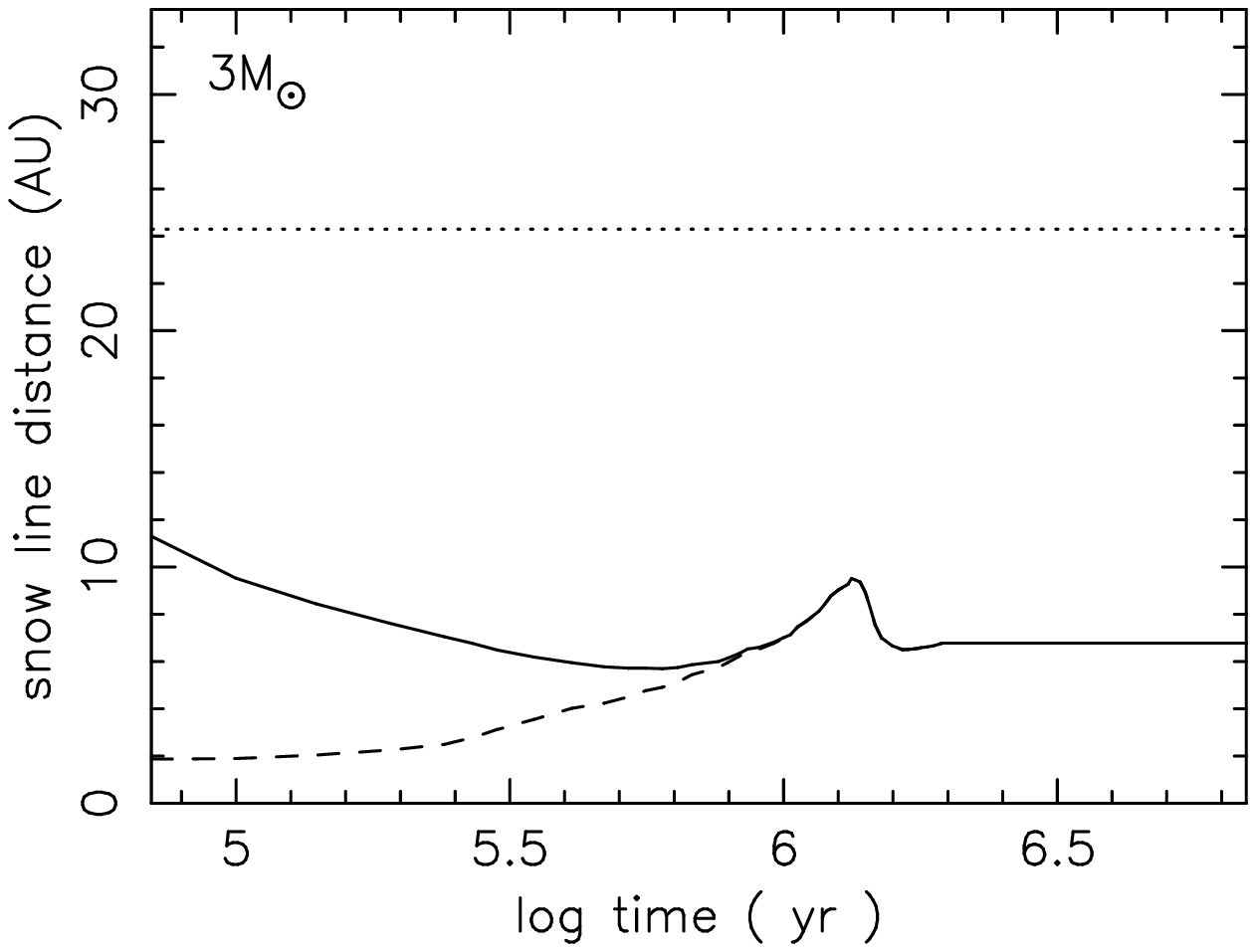}
  \caption{Location of the snow line ($a$ at $T_{mid} = 170$\,K) over
    time for 0.6, 1, 2, and 3\,$M_\odot$ stars (left to right, and
    down) with irradiation only (using \citet{1999ApJ...525..772P} PMS
    tracks, dashed line), and irradiation + accretion (solid
    line). The disks have surface densities $\sigma = \sigma_{\rm
      MMSN} M_\star / M_\odot$. Included for reference is $a_{snow} = 2.7
    M_\star/M_\odot$\,AU as used by \citet{2005ApJ...626.1045I}
    (dotted line).}\label{fig:zones_diff}
\end{figure*}

Figure \ref{fig:zones_diff} shows the location of the snow line in
disks with $\sigma = \sigma_{\rm MMSN} \, M_\star / M_\odot$ for
several different stellar masses over time for irradiation only, and
for accretion + irradiation. We locate the snow line where $T_{mid} =
170$\,K. More detailed derivations of this temperature
\citep[e.g.][]{2004M&PS...39.1859P,2006ApJ...640.1115L} do not change
the snow line distance significantly. For PMS stellar properties we
use \citet{1999ApJ...525..772P} tracks. For comparison, we also show
the (fixed) snow line distance for stars on the main-sequence
\citep[e.g.][]{2005ApJ...626.1045I}.

Looking first at the Solar case, the snow line moves inward over
time. This movement is always determined by viscous accretion, and its
decay over time. Our snow line crosses the ``canonical'' distance of
2.7\,AU at $5 \times 10^5$\,yr. For disks with accretion rates so low
that irradiation dominates, the snow line still moves inward over
time, as illustrated by the dashed line for $T_{irr}$.

For more massive stars, $T_{irr}$ begins to dominate as accretion
subsides and the star quickly evolves to a significantly greater
main-sequence luminosity. Irradiation becomes important at a few Myr
for 2\,$M_\odot$ stars, and $\sim$1\,Myr for 3\,$M_\odot$ stars. The
large discrepancy between our snow line, and that of
\citet{2005ApJ...626.1045I} (who considered $0.2\,M_\odot < M_\star <
1.5\,M_\odot$) arises because theirs is based on the main-sequence
luminosity ($L_\star \propto M_\star^4$) and an optically thin disk
($T_{disk}^4 \propto L_\star \, a^{-2}$). With our model, the snow
line distance is less sensitive to stellar mass, allowing icy
protoplanet formation relatively close ($\sim$5--10\,AU) to the
central star for intermediate mass stars. At these closer distances,
the surface density is higher and formation is faster, making it more
likely that protoplanets massive enough to undergo core accretion will
form. For less massive stars, the snow line is still at a few AU,
where isolation times are relatively long, making it difficult to form
cores before the gas disk is dissipated. Comparison of the snow line
distance with typical disk lifetimes of several Myr leads to an
increasing snow line distance with stellar mass.

With the snow line evolution established, we now describe our model of
protoplanet formation.

\section{Protoplanet Formation Model}\label{sec:model}

The MMSN is a simple model disk for the origin of the Solar System,
and has $\sigma(a) = \sigma_0 \, f_{ice} \, a^{-\delta}$, where the
factor $f_{ice}$ represents a jump in surface density at the snow line
distance $a_{snow}$, and $\delta$ is usually 3/2. To extend this model
to a range of stellar masses requires consideration of how disk mass
varies with stellar mass. Observations indicate $M_{disk} \propto
M_\star$ \citep{2000prpl.conf..559N,2006ApJ...645.1498S}; however,
there is a wide range of disk masses at any given stellar mass. Thus,
to extend the MMSN model to a range of stellar and disk masses, we
adopt the surface density relation
\begin{equation}\label{eq:sigma}
  \sigma(a,t) = \sigma_0 \, \eta \, f_{ice} \, \frac{M\star}{M_\odot}
  a_{\rm AU}^{-\delta} \, ,
\end{equation}
where $\sigma_0 = 10$\,g cm$^{-2}$. The factor $\eta$ changes the disk
mass relative to the star (``relative disk mass''), and is varied to
account for the observed range of disk masses at fixed stellar
mass. Current observations suggest $\eta \sim 0.5$--5 ($M_{disk} =
0.01$--0.1\,$M_\star$); $\eta \sim 10$ is the upper limit for disk
stability ($M_{disk} \sim 0.25\,M_\star$). To provide a smooth
transition from $f_{ice} = 1$ for $a \lesssim a_{snow}$ to $f_{ice} =
3$ for $a \gtrsim a_{snow}$, we set
$f_{ice}=1+(\Delta_{ice}-1)/(1+e^x)$ where $\Delta_{ice}=3$,
$x=(a_{snow}-a)/\Delta a_{snow}$ and $\Delta a_{snow}$ is the radial
distance equivalent to a 5\,K temperature change.

Combined with the local orbital period, the surface density sets the
time to form protoplanets, and sets whether protoplanets form early
enough to accrete gas and become gas giants. We introduce a stellar
mass dependence, so our isolation timescale, based on numerical
simulations by \citet{2004AJ....127..513K,2004ApJ...602L.133K},
becomes
\begin{equation}\label{eq:tiso}
  t_{iso} \propto \left( \eta \sigma \right)^{-1} \,
  a^{3/2} \, M_\star^{-1/2} \, .
\end{equation}
The normalisation of equation (\ref{eq:tiso}) depends on the size of
the small objects
\citep[e.g.][]{2004ApJ...614..497G,2006ApJ...652L.133C}. We use
10$^5$\,yr for $\sigma = 10$\,g cm$^{-2}$ at 5\,AU, based on the
likelihood of small fragmented bodies to accrete
\citep[e.g.][]{2004AJ....127..513K}, and the consequent short growth
times
\citep{2004AJ....128.1348R,2006ApJ...652L.133C}. \citet{2005Icar..179..415H}
infer $t_{iso} \lesssim$5$\times 10^5$\,yr for much larger 100\,km
planetesimals. As long as the Jovian timescale is somewhat shorter
than the gas disk lifetime, this choice affects our results little.

Under the assumption that stars all form disks in $\sim$10$^5$\,yr
\citep[based on an infall rate of $10^{-5}\,M_\odot$
yr$^{-1}$,][]{1999ApJ...525..772P}, and that planetesimal formation is
relatively fast \citep{2000SSRv...92..295W,2005A&A...434..971D}, we
add a constant offset of $10^5$\,yr to the isolation timescale to
reconcile the timing of isolation with disk and stellar
evolution. Though this time is uncertain, removing or moderately
modifying the offset does not affect our results significantly because
$t_{iso}$ is usually $\gtrsim$10$^5$\,yr.

We adopt a range of disk masses, integrated from the inner disk radius
to 60\,AU, with a gas to solids ratio of 100. For $\delta = 3/2$,
$\eta = 4$ corresponds to a relatively massive disk $M_{disk} =
0.1\,M_\star$. This enhancement is our baseline model and yields the
surface density and core mass needed to form Jupiter on reasonable
time scales \citep{1996Icar..124...62P,2000ApJ...537.1013I}. For
$\delta = 1$, smaller $\eta$ yields the same disk mass because more
mass is placed at larger radii. For $\eta = 1$ and $\delta = 1$,
$M_{disk} = 0.12\,M_\star$.

\section{Regions that form Gas Giant Cores}\label{sec:results}

The successful formation of a gas giant planet by core accretion
requires satisfaction of two main conditions. A core must form while
the gaseous component of the circumstellar disk is still present, and
it must be massive enough to attract a large atmosphere before this
gas is dispersed. Prior to isolation, accreted planetesimals and a
sub-critical protoplanet mass limit gas accretion. After isolation, if
a protoplanet is massive enough, and forms while the gas disk is still
present, significant gas accretion proceeds. This separation into two
classes, gas giants and ``failed cores,'' reflects the expected
paucity of 20--100\,$M_\oplus$ planets over a range of stellar masses
\citep{2005ApJ...626.1045I}.

To form a gas giant in less than 10$^7$\,yr, various studies suggest a
minimum core mass of 5--10\,$M_\oplus$
\citep{2000ApJ...537.1013I,2003Icar..166...46I,2005Icar..179..415H}. Although
the core mass depends on the planetesimal accretion rate and the
opacity, the derived sensitivity is weak \citep[$M_{core} \propto
\dot{M}^{0.25}$;][see also
\citet{2006ApJ...648..666R}]{2000ApJ...537.1013I}. A limited reservoir
of planetesimals to accrete after isolation means high accretion rates
cannot be sustained (and $\dot{M}$ will decrease), while low accretion
rates lower $M_{core}$. Thus, we adopt a minimum core mass $M_{core} =
10\,M_\oplus$ as a baseline, and consider $M_{core} = 5\,M_\oplus$ in
\S \ref{sec:varying}.

The timescale for gas dissipation sets our second restriction. The
gaseous component of the disk disperses in $\lesssim$10\,Myr
\citep{1995Natur.373..494Z,2006ApJ...651.1177P}. With $\dot{M} \propto
\sigma$, the dissipation timescale for a viscous disk $t_d \propto
M_{disk} / \dot{M} \sim {\rm constant}$ for our assumptions. Because the
disk mass decreases significantly ($\sim$60\%) in 1\,Myr, we adopt
$t_{core} = 1$\,Myr as a typical maximum core formation time for all
disks. Henceforth we reserve the word ``core'' for a protoplanet with
$M_{iso} > M_{core}$ and $t_{iso} < t_{core}$. Relatively little is
known about the evolution of the gaseous component of the disk; we
comment further on the consequences of varying $t_{core}$ and other
parameters in \S \ref{sec:varying}.

To investigate locations within circumstellar disks where gas giant
cores form, we first derive results for a $1\,M_\odot$ star, and then
consider a range of stellar masses. We restrict our study to stars
with masses 0.2--4\,$M_\odot$. For stars with masses $<$0.2\,$M_\odot$
our model does not form gas giants. The short main-sequence lifetime
of massive stars ($M_\star > 4\,M_\odot$) makes them much less likely
targets for planet detection. Some oligarchs do not reach masses
sufficient for core accretion. Those cores that do form compete with
other cores for dynamical space in the disk. We defer consideration of
these objects to \S \ref{sec:other}.

\subsection{The Solar Example}\label{sec:MMSNsolar}

In this subsection we show how the MMSN disk model, the moving snow
line, and the isolation mass and time combine to give a picture of the
Solar System structure at $\sim$1\,Myr.

\begin{figure}
  \plotone{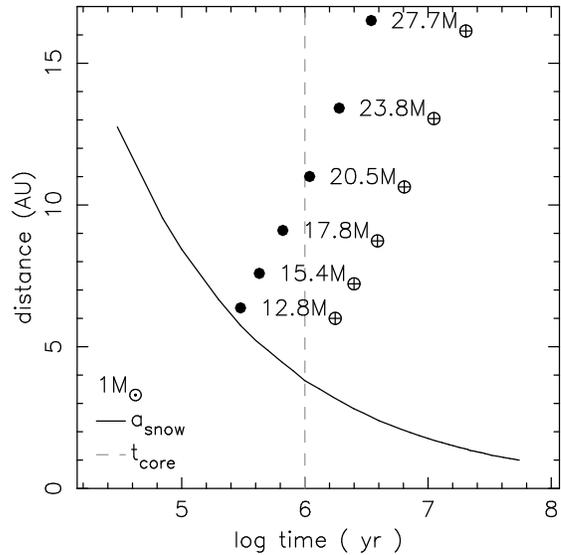}
  \caption{Isolation mass (filled circles, labelled with $M_{iso}$) as
    a function of radial distance and PMS model time, for a solar mass
    star with the MMSN model with $\eta=4$ and $\delta = 3/2$. Masses
    are spaced at $8\,R_H$ intervals and only shown outside the snow
    line. The solid line shows $a_{snow}$ over time, and the dashed
    vertical line is the time $t_{core} = 1$\,Myr.}\label{fig:pfzones}
\end{figure}

Figure \ref{fig:pfzones} shows isolation masses for the MMSN model
beyond the snow line with $\eta=4$, as a function of time and radial
distance from the Sun. The isolation mass and timescale are calculated
from the equations described in \S \ref{sec:model}. Accretion and PMS
tracks from \citet{1999ApJ...525..772P} set $a_{snow}$ as described in
\S \ref{sec:motivation}.

When the first objects reach isolation, the Sun is in the early stages
of its PMS contraction, and the accretion rate is
$\sim$10$^{-7}\,M_\odot$ yr$^{-1}$. Consequently the snow line is at
$\sim$6\,AU, which determines where the innermost icy protoplanet
forms. This protoplanet is massive enough to become a gas giant (a
core), so we refer to this position as the inner edge of the
core-forming region. In the absence of significant migration from disk
interaction, this result may help explain why Jupiter is at 5\,AU. As
the Sun continues to contract and accretion decreases, isolation is
reached at ever increasing distances beyond the snow line. Eventually,
the isolation time becomes longer than $t_{core}$, and protoplanets
form too late to undergo core accretion. Isolation masses increase
with distance from the Sun, so $t_{core}$ always sets the outer edge
of the core-forming region, and the number of cores that form. If the
cores are spaced by 8\,$R_H$ at isolation (as in Figure
\ref{fig:pfzones}), then $\sim$4 cores form in this region and the
region extends from $\sim$6--11\,AU, similar to the region containing
Jupiter and Saturn today.

\subsection{A Range of Stellar Masses}

We now consider how the core-forming region for a solar mass star
changes with stellar mass between $0.2\,M_\odot$ and $4\,M_\odot$. The
processes described in \S \ref{sec:MMSNsolar} still apply, but
differences arise due to the linear relation between disk surface
density and stellar mass, the different orbital periods around other
stars, and the changing snow line distance due to the evolution of
accretion and the central star.

\begin{figure*}
  \plottwo{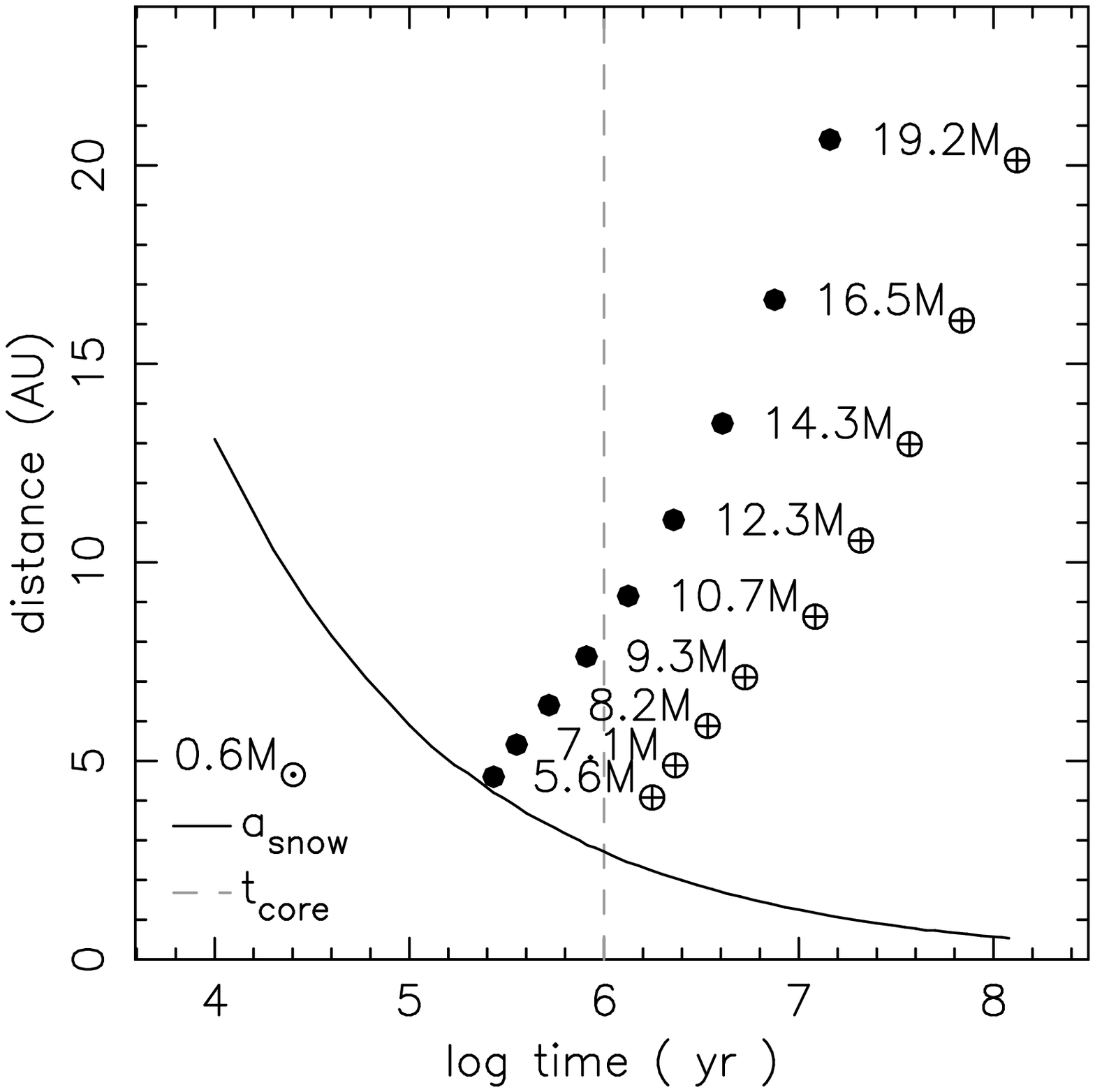}{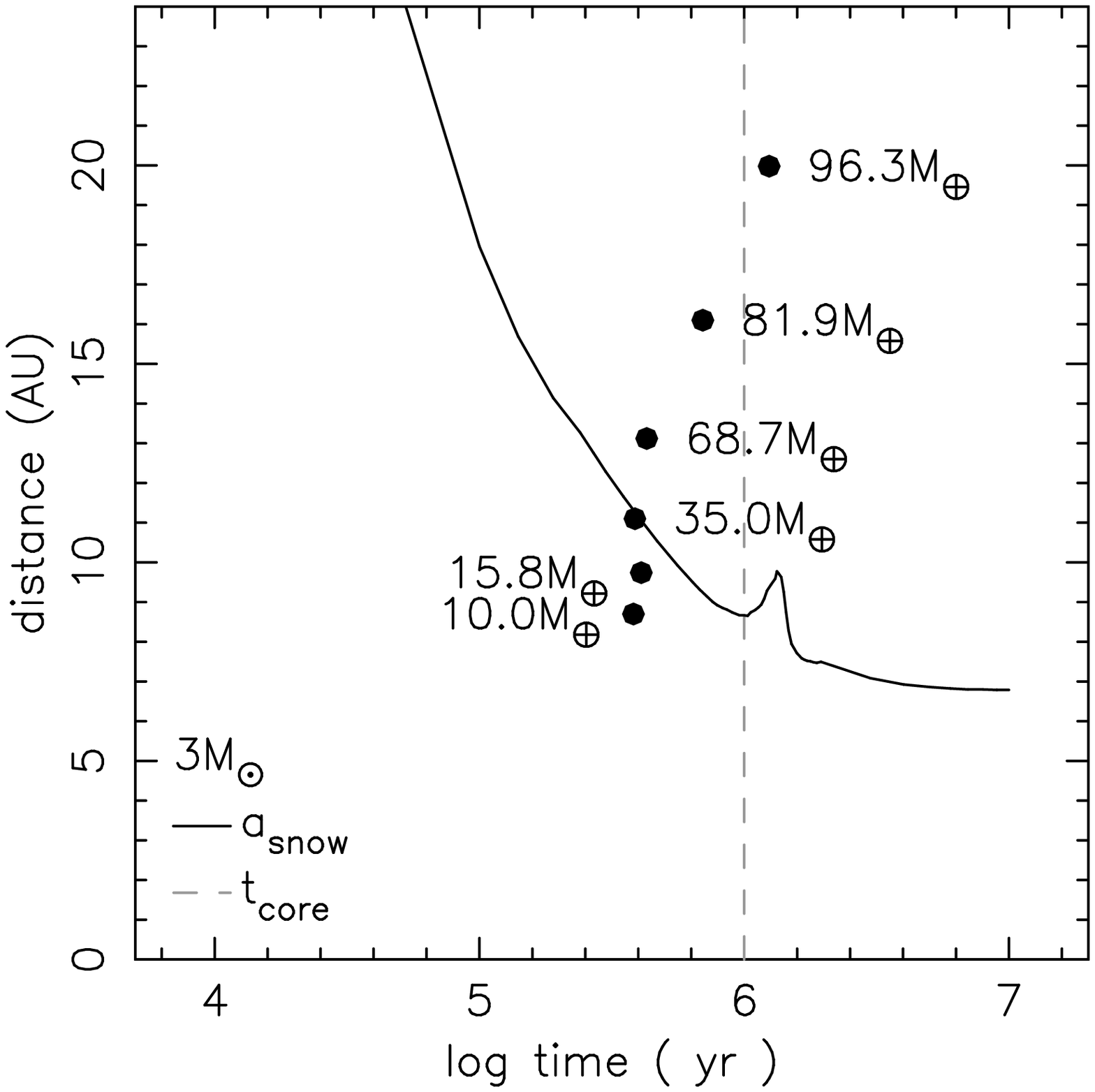}
  \caption{Same as Figure \ref{fig:pfzones}, but for 0.6\,$M_\odot$
    (left) and 3\,$M_\odot$ (right). Isolation masses are only plotted
    outside the snow line, or where $M_{iso} >
    M_{core}$.}\label{fig:pfzones1}
\end{figure*}

Figure \ref{fig:pfzones1} shows isolation masses for 0.6 and
3\,$M_\odot$ stars, with $t_{core} = 1$\,Myr and the same relative
disk mass as Figure \ref{fig:pfzones}. The lower mass star does not
form any cores, as the first object with $M_{iso} > M_{core}$ forms
after $t_{core}$. However, large 5--10\,$M_\oplus$ objects still form
(see \S \ref{sec:failed}). A longer $t_{core}$ allows some of these to
become cores, so whether lower mass stars form gas giants is sensitive
to both $M_{core}$ and $t_{core}$.

The 3\,$M_\odot$ star forms its innermost core just inside the snow
line at $\sim$8\,AU, and the outermost is at $\sim$20\,AU. Cores can form
before $t_{core}$ at greater distances due to decreased $P$ and
increased $\sigma$. The greater surface density in disks around these
stars allows rocky cores to form interior to the snow line \citep[see
also][]{2005ApJ...626.1045I}. The large mass of the cores
($\gg$$M_{core}$) in these relatively massive disks probably allows
some gas accretion prior to isolation, which requires a numerical
model of growth for more investigation (see also \S
\ref{sec:varying}).

\begin{figure*}
  \plotone{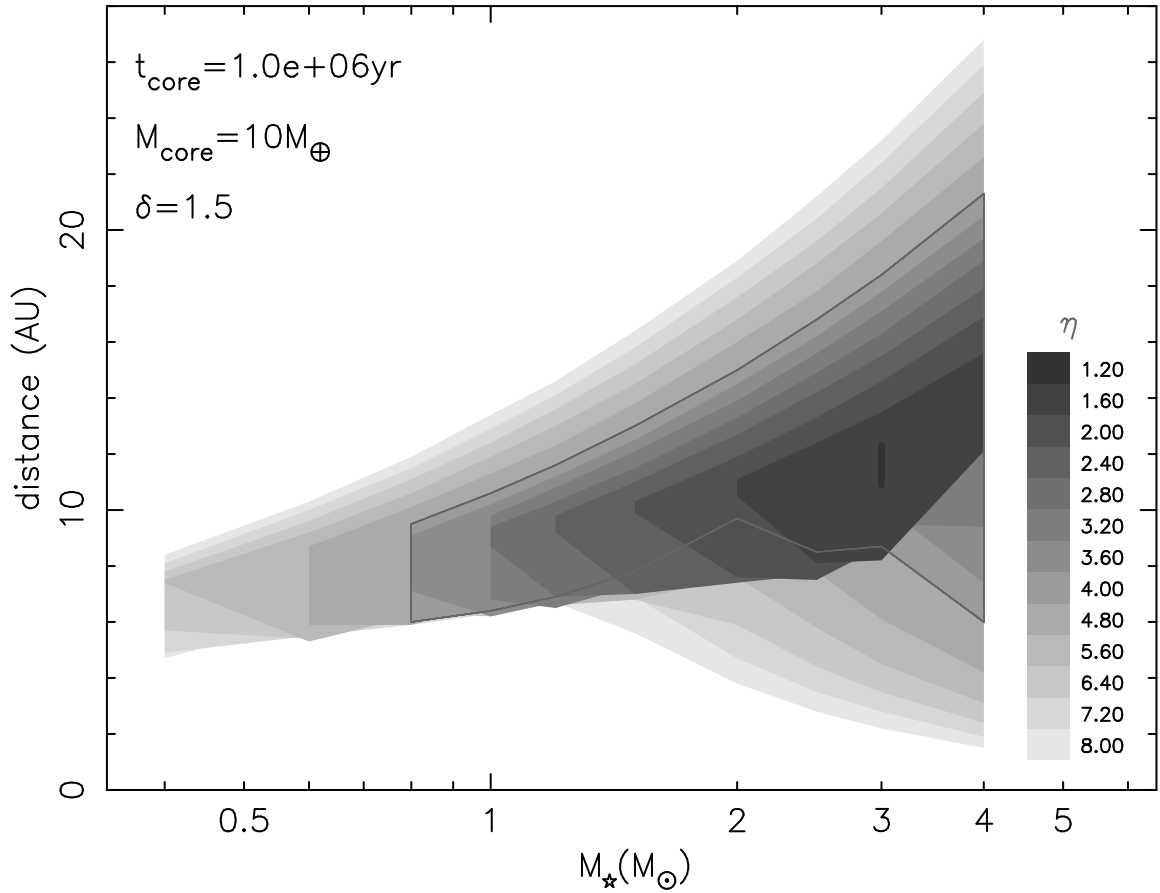}
  \caption{Regions where $10\,M_\oplus$ cores form in $\leq$$10^6$\,yr
    as a function of radial distance and stellar mass. Each contour
    represents the inner, outer, and stellar mass limits for a
    particular $\eta$ (and corresponding $M_{disk}$), as
    shown by the legend. Our baseline model, $\eta=4$ is outlined. For
    $\eta = 1.2$, only $3\,M_\odot$ stars form
    cores.}\label{fig:cores_a_ms1}
\end{figure*}

The results over a range of stellar masses can be combined into a
single figure that considers the core-forming regions as a function of
stellar mass. Figure \ref{fig:cores_a_ms1} shows the core-forming
region as a function of radial distance and stellar mass for our
model, with standard parameters of $t_{core}=1$\,Myr,
$M_{core}=10\,M_\oplus$, and $\delta=3/2$. Each contour represents the
inner, outer, and stellar mass limits for forming cores with a
particular relative disk mass. On the $\eta = 4$ contour (outlined in
the figure), the 6--10\,AU range from Figure \ref{fig:pfzones}
contributes the points $y=6$\,AU and $y=10$\,AU for $x =
1\,M_\odot$. Similarly, the 8--20\,AU range from the right panel of
Figure \ref{fig:pfzones1} contributes points at $x = 3\,M_\odot$.

In general, Figure \ref{fig:cores_a_ms1} shows that as stellar mass
increases, disks with lower relative disk masses form cores,
and the width of the regions where these cores form increases. Stars
more massive than $\sim$1.2\,$M_\odot$ form rocky cores interior to
the snow line for relatively high disk masses. The core-forming region
expands outward with increasing relative disk mass because the
isolation timescale becomes shorter. Doubling the disk mass allows $a$
to increase by a factor of $\sim$1.6 to keep the same $t_{iso}$
(eq. \ref{eq:tiso}).

As stellar mass increases, cores form in disks with decreasing
relative disk mass. As $\eta$ decreases, the inner edge of the
core-forming region moves inward, because the accretion rate is lower
and the disk has lower optical depth (so the mid-plane is cooler), and
the snow line has evolved closer to the star by the time isolation is
reached. The snow line is roughly the lower edge of the darker (lower
$\eta$) contours.

If $M_{iso}$ does not jump to a value $>$M$_{core}$ due to $f_{ice}$
at the snow line, a core still forms further out. Thus, the inner edge
moves to greater distances as the stellar mass---and hence absolute
disk mass---decreases for fixed relative disk mass. However, the
$t_{core}$ restriction means that oligarchs around sufficiently
low-mass stars do not reach isolation in time, and that all contours
have a lower stellar mass limit.

The lowest disk mass that forms cores has $\eta = 1.2$, and only does
so for 3\,$M_\odot$. For more massive stars, irradiation overcomes
accretion as the star reaches the main-sequence, and the larger snow
line distance makes formation of cores more difficult \citep[see
also][]{2005ApJ...626.1045I}. Thus 3\,$M_\odot$ stars are the most
likely to form at least one gas giant, as they form cores over the
widest range of disk masses.

The width of the regions where cores form increases with stellar mass
\citep[Figure \ref{fig:cores_a_ms1},][]{2005ApJ...626.1045I}. The
spacing of cores remains roughly constant with different stellar mass
($M_{iso} \propto \sigma^{3/2} / \sqrt{M_\star} \propto M_\star$ and
$R_H \propto (M_{iso}/M_\star)^{1/3}$ for $\delta=3/2$, see also
Figs. \ref{fig:pfzones} \& \ref{fig:pfzones1}). However, the number of
cores is not linearly related to the region width, since the spacing
becomes wider with increasing distance. The width of the regions
depends strongly on disk mass, particularly for disks that form cores
interior to the snow line. The increasing width of the core-forming
regions suggests that the number of cores (and therefore planets) in
individual planetary systems increases with stellar mass.

To summarise, the range of relative disk masses that form gas giant
cores increases with stellar mass, as does the width of the regions
they form in. The first result leads to the expectation that the
likelihood of forming gas giants increases with stellar mass. We make
a quantitative prediction in \S \ref{sec:probability}.

While the core-forming regions are our primary interest, there are
large regions of parameter space where oligarchs are relatively
massive, but will not form gas giants. We consider these planets now.

\subsection{Other Planets}\label{sec:other}

The discovery of exoplanets with masses smaller than Neptune suggests
that planet formation might often yield failed cores---oligarchs that
did not accrete gas from the disk. Neptune and Uranus may be
considered failed cores. Several theoretical studies \citep[e.g.][this
paper]{2004ApJ...612L..73L,2005ApJ...626.1045I} suggest that failed
cores are more common around low mass stars.

\subsubsection{Failed Cores}\label{sec:failed}

In our model, there are three ways to produce failed cores. Objects
that form too late ($t_{iso} > t_{core}$), or form with insufficient
mass to accrete gas ($M_{iso} < M_{core}$) are failed cores. Although
all cores within the core-forming region can potentially accrete gas,
dynamical interactions among the cores may eject one or more into
regions with a small gas surface density
\citep[e.g.][]{1999Natur.402..635T} or from the system entirely
\citep[e.g.][]{2004ApJ...614..497G,2007ApJ...661..602F}. This
mechanism occurs in a random (and currently unquantifiable) fraction
of models with $M_{iso} > M_{core}$ and $t_{iso} < t_{core}$. In our
model failed cores are more common around low-mass stars because
isolation masses are smaller, and isolation times are longer.

Apparent failed cores may also form by collisions over long timescales
\citep{2006ApJ...650L.139K}. The mass of these icy planets may be
limited by the likelihood of collisions vs. ejections during the final
stages of coalescence \citep{2004ApJ...614..497G}.

\subsubsection{Ocean Planets}

The diversity of observed extra-solar planets led
\citet{2004Icar..169..499L} to suggest that 1--$10\,M_\oplus$ icy
planets that form in the region beyond the snow line may migrate
inward to $\sim$1\,AU, where the outer layers subsequently ``melt.''
With masses too low to accrete much gas, these planets are less dense
than a rocky planet of equivalent mass and harbour deep oceans. Hence
\citeauthor{2004Icar..169..499L} call these ``ocean planets.''

The increase in luminosity of stars with masses
$\gtrsim$2.5\,$M_\odot$ as they reach the main-sequence provides an
alternative \emph{in situ} formation mechanism for ocean planets, as
they may have insufficient mass for significant migration. We outline
the concept briefly, because these planets are difficult to detect. At
times $\gtrsim$1--10\,Myr, the snow line moves to $\sim$10\,AU as the
disk becomes optically thin.\footnote{The snow line has less meaning
  at these times, since there is little gas to condense into ices. The
  equilibrium temperature of objects is a more relevant concept.}
Failed cores in the range 1--$10\,M_\oplus$ can therefore achieve
their final mass outside the snow line in $\sim$1\,Myr, and without
migrating, later find themselves in a much warmer region when the star
reaches the main-sequence.

Though all stars more massive than the Sun undergo an increase in
luminosity as they settle onto the main-sequence
\citep[e.g.][]{1999ApJ...525..772P}, the temperature at the early snow
line distance of $\sim$7\,AU must increase enough to melt ice and
maintain oceans. For 2.5\,$M_\odot$, there is $\sim$1\,AU overlap
between the early snow line, and the final habitable zone
distance---at an equilibrium temperature of $\sim$245\,K
\citep{1993Icar..101..108K}---with room for a few cores that form
\emph{in situ} just beyond the snow line.

\subsection{Sensitivity to Model Assumptions}\label{sec:varying}

Our model is simplified, but captures some important concepts. In this
section, we show that our results remain for realistic variations on
our model assumptions.

We use a simple model for the temperature profile of an irradiated,
accreting, flared disk, which sets the location of the snow line. As
shown in Figure \ref{fig:cores_a_ms1}, $\sim$AU changes in the snow
line distance affects where the innermost cores originate, but there
is little change in the range of disk masses that forms massive cores
for a given stellar mass.

\begin{figure}
  \plotone{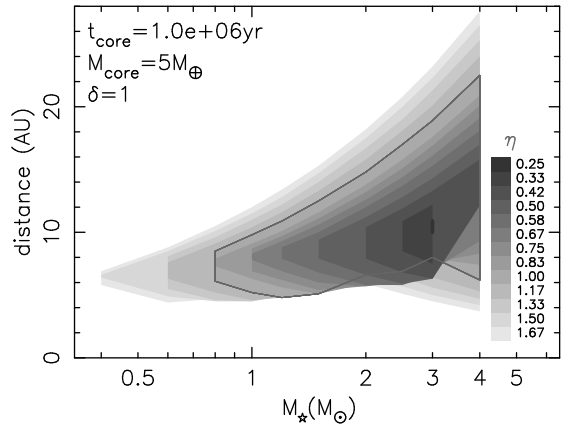}
  \caption{Same as Figure \ref{fig:cores_a_ms1}, but for $\delta = 1$
    and $M_{core} = 5\,M_\odot$. For $\eta = 0.25$, only $3\,M_\odot$
    stars form cores. Though the $\delta = 1$ disk has different
    $\eta$, it covers the same range of disk masses as Figure
    \ref{fig:cores_a_ms1} with $\delta = 3/2$, with the exception of
    the lowest disk mass ($\eta=1.2$) from that
    figure.}\label{fig:cores_a_ms2}
\end{figure}

The disk surface density profile is uncertain: the MMSN assumes the
Solar System planets formed \emph{in situ}. In the standard MMSN
model, $\delta = 3/2$ and $\eta=4$ yield the surface density needed to
form a massive core near Jupiter. If $\delta = 1$ and $\eta = 1$ ,
then $M_{iso} \approx 5$ (13)\,$M_\oplus$ at 5 (10)\,AU, which are
similar to the inferred core masses for Jupiter and Saturn
\citep{2004ApJ...609.1170S}. Figure \ref{fig:cores_a_ms2} shows the
core-forming regions for $\eta = 1$, $\delta = 1$ and $M_{core} =
5\,M_\oplus$. The lower $\sigma$ in the core-forming regions needed to
keep the same disk mass makes it harder to form cores with $M_{core} =
10\,M_\oplus$, but with $M_{core} = 5\,M_\oplus$ the regions are
similar to our baseline model.

\begin{figure*}
  \plottwo{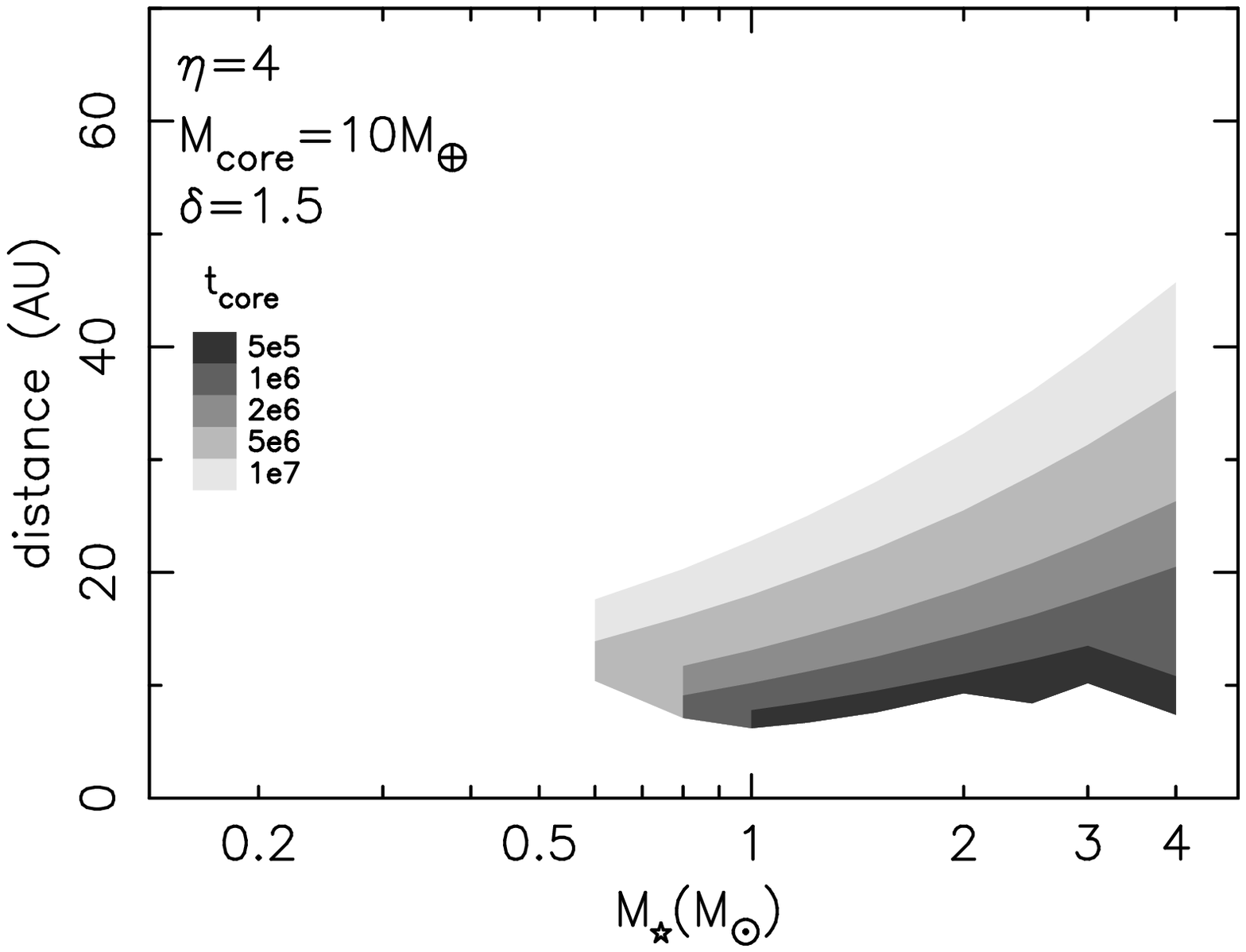}{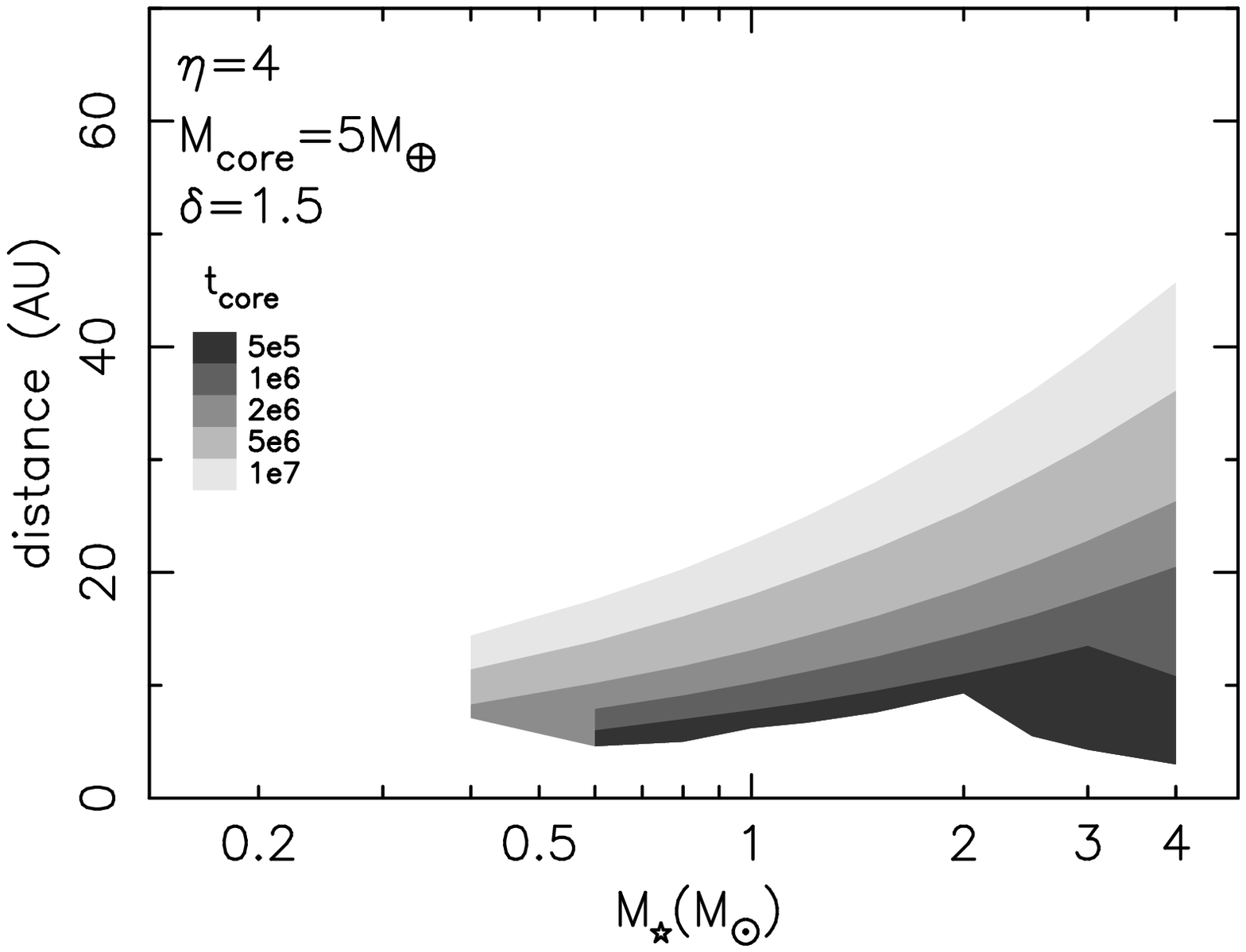}
  \vspace{10pt}\\
  \plottwo{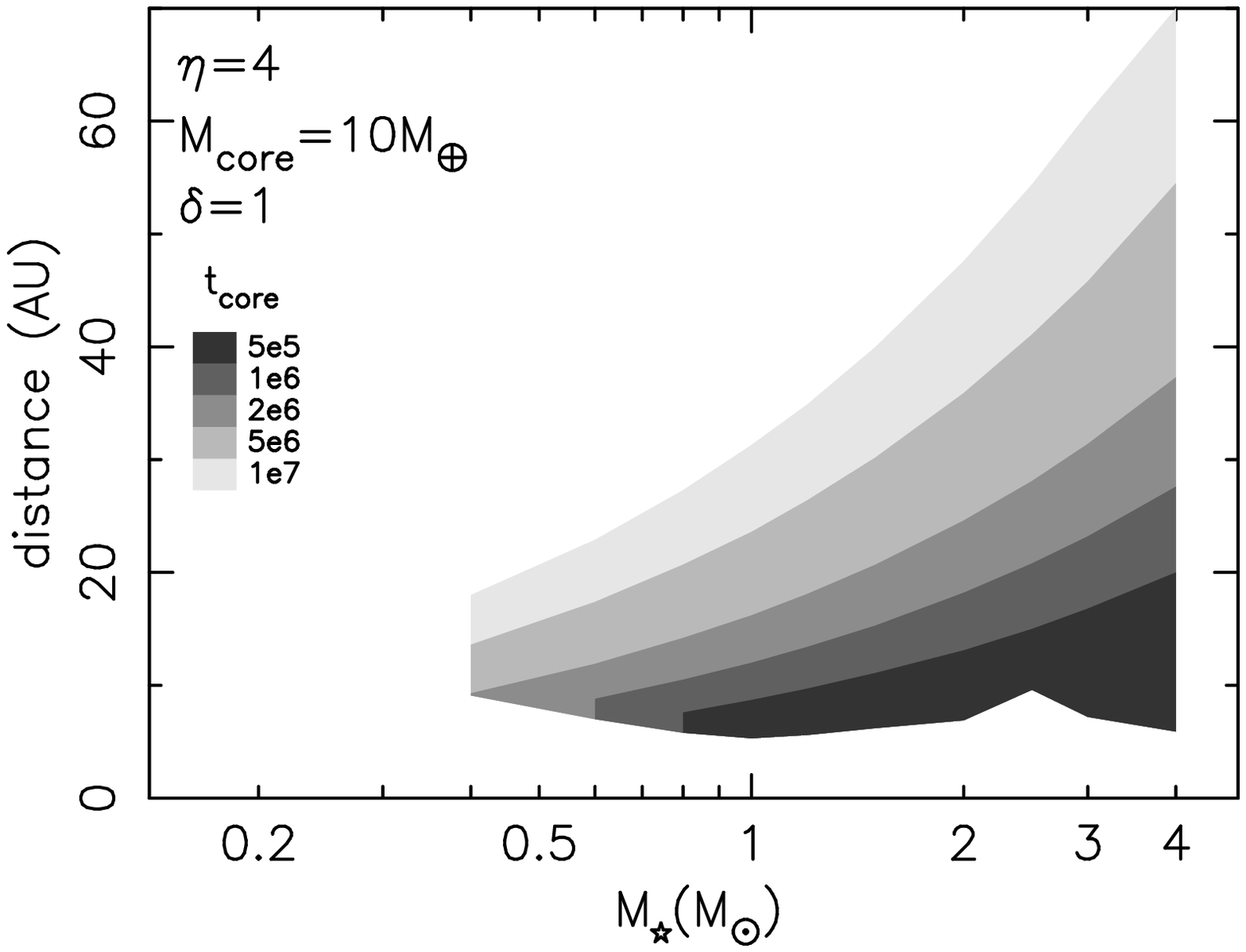}{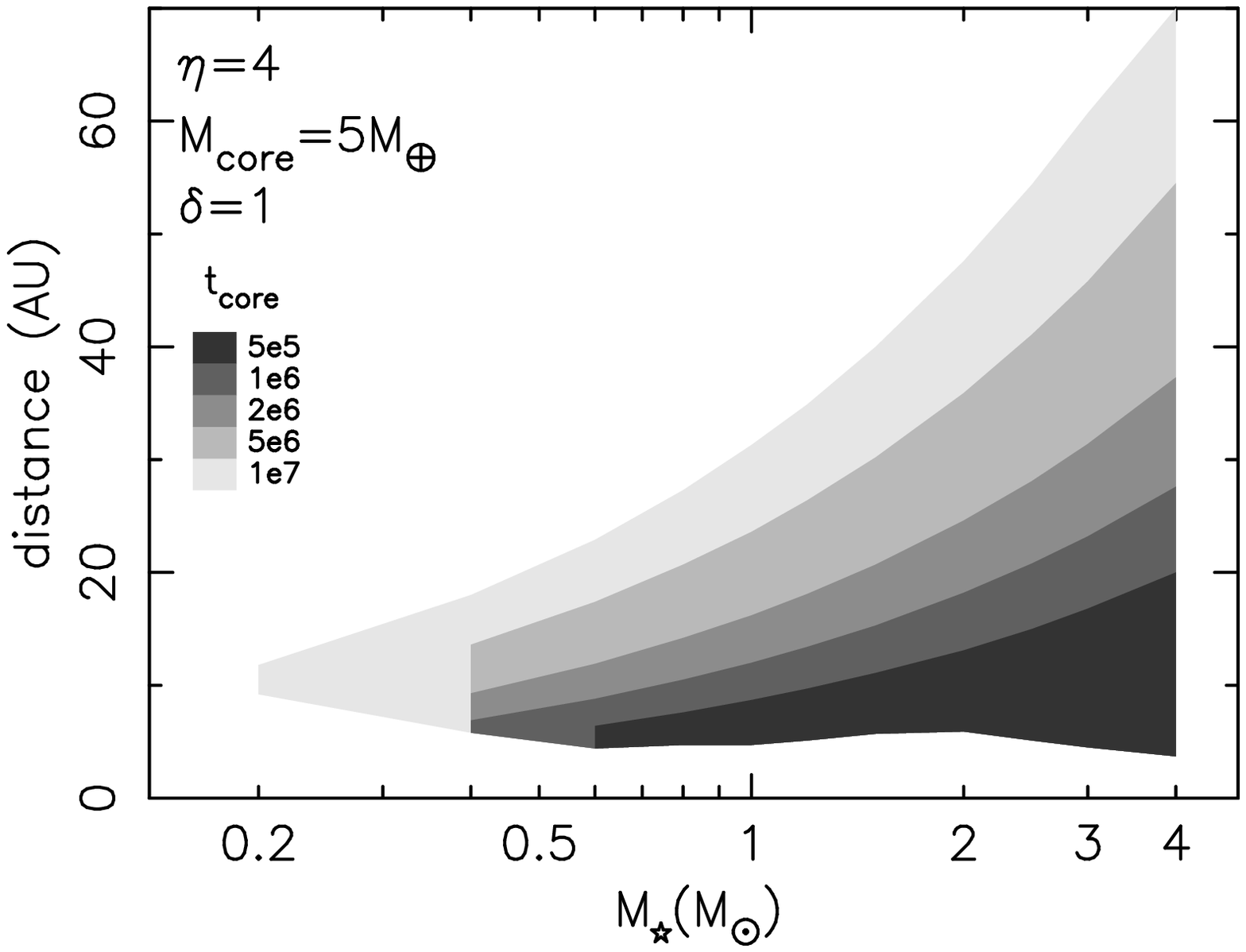}
  \caption{Similar to Figure \ref{fig:cores_a_ms1}, but for fixed
    $\eta = 4$. Contours represent different $t_{core}$ as
    indicated by legends. Top (bottom) panels are $\delta = 1.5$
    ($\delta = 1$), and left (right) panels are $M_{core} =
    10\,M_\oplus$ ($M_{core} =
    5\,M_\oplus$).}\label{fig:cores_a_ms_tc}
\end{figure*}

There is little observational constraint of gaseous inner disk
lifetimes, so the least certain of the parameters we specify is
$t_{core}$. The isolation time is $t_{iso} \propto a^3$ (when $\delta
= 3/2$), so doubling $t_{core}$ allows the outer edge to move outward
by a factor of about 1.3. Figure \ref{fig:cores_a_ms_tc} shows how
changing $t_{core}$ alters the core-forming region with $\eta = 4$ for
$M_{core} = 5\,M_\oplus$ and 10\,$M_\oplus$, and $\delta = 1$ and
3/2. Longer gaseous disk lifetimes lead to more gas giant cores. The
general trend is to extend the regions to lower stellar masses and to
greater radial distances. Cores that take longer to form---due to
smaller $P$ or $\sigma$---can reach isolation before the gas disk is
dissipated.

Dust disks around spectral types earlier than $\sim$G have somewhat
shorter lifetimes than disks around lower mass stars
\citep{2001AJ....121.2065H,2007ApJ...659..599C}. If gas is removed on
timescales similar to infra-red excesses for a range of spectral
types, $t_{core}$ is shorter and the outer edge of the core-forming
region for these stars moves in (Fig.
\ref{fig:cores_a_ms_tc}). Observations of evolved stars can test
whether a strongly stellar mass dependent $t_{core}$ (or some other
process) defines an upper stellar mass limit for gas giant formation.

The short timescale for disk removal by the central star---inferred
from the lack of transition disks
\citep[e.g.][]{1995ApJS..101..117K,2001MNRAS.328..485C,2006MNRAS.369..229A}---and
the likely large radial distance in the disk for external influence by
massive stars \citep{2004ApJ...611..360A}, supports our assumption
that the location of photoevaporation is unimportant. Although the
expected distance for external photoevaporation in ``typical''
clusters reaches closer to the central star for low-mass stars
\citep{2004ApJ...611..360A}, it lies outside the core-forming region
for our standard case of $M_{core} = 10\,M_\oplus$, $t_{core} =
10^6$\,yr, and $\delta = 1.5$. The core-forming and external
photoevaporation regions begin to overlap for $M_\star \lesssim
0.5\,M_\odot$ and $t_{core} \gtrsim 5 \times 10^6$\,yr.

The size of planetesimals is uncertain, as is the isolation time that
results from their accretion by oligarchs. Our choice of $10^5$\,yr
for the Jovian core is relatively short, and is based on likely
fragmentation \citep[e.g.][]{2004AJ....127..513K} and the rapid
accretion of small $\sim$100\,m planetesimals in the shear dominated
regime \citep[e.g.][]{2004AJ....128.1348R,2006ApJ...652L.133C}. If
planetesimals are larger and $t_{iso}$ is longer (e.g. 1\,Myr),
similar results can be obtained by simply using a longer (yet still
reasonable) $t_{core} \sim 3$\,Myr.

Choosing $M_{core} = 5\,M_\oplus$ allows core formation in less
massive disks. Halving $M_{core}$ allows a disk 1.6 times ($M_{iso}
\propto \sigma^{3/2}$) less massive to form cores in the same region,
and also extends the region to lower stellar masses
(Fig. \ref{fig:cores_a_ms_tc}).

We have chosen to ignore type I migration, where linear theory
predicts that protoplanets excite spiral density waves in the gas
disk, and migrate toward the central star on 0.01--0.1\,Myr timescales
\citep{2002ApJ...565.1257T}. Recent studies indicate that for cores
less massive than $\sim$10\,$M_\oplus$ the timescale is longer
\citep{2006ApJ...652..730M}, and may be reduced to a random walk due
to magnetohydrodynamic turbulence \citep{2004MNRAS.350..849N}. For
cores with masses $\gg$10\,$M_\oplus$ (Fig. \ref{fig:pfzones1}) in
relatively massive disks around intermediate mass stars, core
accretion will likely occur before isolation, while planetesimals are
still being accreted \citep{2006ApJ...648..666R}. The successful
formation of a gas giant then depends on whether the planet can reach
a gap-opening mass before migrating into the central star.

Once a planet opens a gap in the disk it has survived the type I
migration regime, but its continued existence is not
guaranteed. Depending on the disk viscosity and lifetime, the planet
can still migrate onto the central star by type II migration.

In summary, the simplicity of our model means that reasonable changes
in the input parameters change the results little. Future development
of the model can include a more complete treatment of more complicated
physical processes.

\section{Discussion}\label{sec:discussion}

The age of direct planet detection is approaching (e.g. NICI Campaign
on Gemini South), where discoveries will be pushed to larger
semi-major axes. In addition, radial velocity surveys now extend over
a wider range of stellar masses
\citep[e.g.][]{2002ApJ...576..478F,2006PASP..118.1685B,2007arXiv0707.2409J,2007arXiv0704.2455J}. Our
goal is to develop a theory of planet formation that extends over the
observational range to make testable predictions, and to develop
greater insight into the processes that produce the observed diversity
of planetary systems.

For planets orbiting giant stars, there is a downward shift in the
planet-metallicity distribution by $\sim$0.3 dex
\citep{2007arXiv0707.0788P}. This result is not surprising in the
context of our model. In Figure \ref{fig:cores_a_ms1}, the lowest
relative disk mass that forms cores roughly halves from 1 to
2\,$M_\odot$, which corresponds to a $-$0.3\,dex change in
metallicity. Thus, we naturally expect the lower end of the
metallicity distribution of higher mass stars to be shifted. However,
we do not expect the high metallicity end of the distribution to move,
since these disks can still form cores.

There are now sufficient planet discoveries to start quantifying
trends across a range of stellar masses, which allows the first steps
towards comparison with planet formation theories that consider the
mass of the central star \citep[e.g.][]{2005ApJ...626.1045I}. Though
sample numbers are small, studies of $\gtrsim$1.3\,$M_\odot$ giants
indicate that giant planet frequency increases with stellar mass in
the range 0.1--2\,$M_\odot$ \citep{2007arXiv0707.2409J}. We now
calculate what our model predicts for the probability of forming gas
giants as a function of stellar mass.

\subsection{Gas Giant Frequency and Stellar Mass}\label{sec:probability}

Assuming all stars are born with a distribution of disk masses, we can
estimate the probability $P_{GG}$ of a star forming at least one gas
giant as a function of stellar mass. Though comparison with observed
disk masses is uncertain, we follow \citet{2005ApJ...626.1045I} and
adopt a Gaussian distribution in terms of $x = \log M_{disk}/M_\star$,
where $P_{disk} \propto \exp \left( - \left( x - \mu \right)^2 / 2
  \sigma_{ln}^2 \right)$ with standard deviation $\sigma_{ln} = 1/3$,
centered on $M_{disk} = 0.03\,M_\star$ (e.g. $\mu \approx -1.5$). This
distribution is similar to data compiled by
\citet{2000prpl.conf..559N}, which is sensitive to all disk masses
that form a core in our baseline model. Because our model is based on
parameters that change with stellar mass (such as isolation time and
disk mass), the relative probability of forming gas giants is our main
concern. Effects that may set the absolute probability, such as
survival of migrating planets, are not included. To make contact with
observations, we therefore normalise our results to 6\% for Solar-mass
stars \citep{2007prpl.conf..685U}.

\begin{figure}
  \plotone{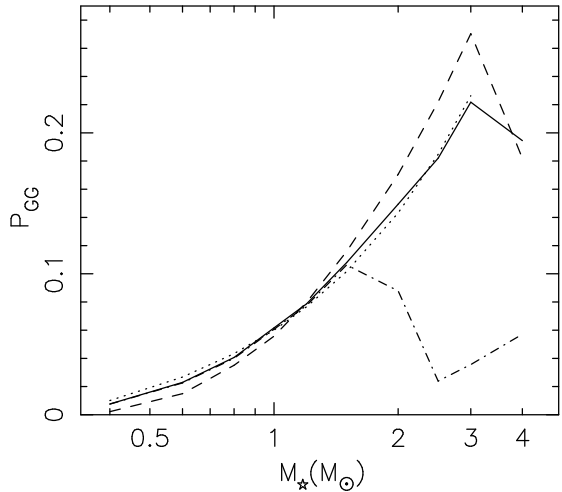}
  \caption{Probability of a star harbouring at least one gas giant
    planet as a function of stellar mass for our baseline model (solid
    line), and $\delta = 1$ and $M_{core} = 5\,M_\oplus$ (dashed
    line). The thin dotted line is a fitted line of constant slope
    $P_{M_\star} = 0.20 \, M_\star - 0.06$. The dot-dashed line has
    $a_{snow} \propto 2.7 M_\star^2$\,AU, for comparison with
    \citet{2005ApJ...626.1045I}. All curves are normalised to 6\% at
    1\,$M_\odot$ via a straight line fit.}\label{fig:cores_f_ms}
\end{figure}

Figure \ref{fig:cores_f_ms} shows the likelihood of a star harbouring
at least one gas giant planet as a function of stellar mass for our
baseline model (Fig.\ref{fig:cores_a_ms1}) and a model with $\delta =
1$ and $M_{core} = 5\,M_\oplus$ (Fig.\ref{fig:cores_a_ms2}),
normalised to 6\% at 1\,$M_\odot$. To illustrate the difference
between a static, main-sequence scaled snow line, and our evolving
one, we include a model with $a_{snow} \propto 2.7 M_\star^2$\,AU,
similar to the main model of \citet{2005ApJ...626.1045I}. Each point
is the probability a star has a disk in the range that forms cores in
Figure \ref{fig:cores_a_ms1}. This plot assumes that if one or more
cores form, at least one will result in a gas giant. For comparison to
current observations, it also assumes that as they accrete gas, cores
generally migrate or scatter a stellar-mass independent fraction to
observable distances, where the 6\% normalisation applies. Like our
baseline model, different $M_{core}$, $t_{core}$, and $\delta$ have
lines of approximately constant slope ($P_{GG} = m \, M_\star - c$) up
to $\sim$3\,$M_\odot$. Our baseline model has $m = 0.20$ and $c =
0.06$; very few gas giants form by core accretion below
0.3\,$M_\odot$.

With the normalisation, our baseline model predicts 1\% of
0.4\,$M_\odot$ stars, and 10\% of 1.5\,$M_\odot$ stars will
harbour at least one gas giant. Increasing $\sigma_{ln}$ decreases the
range of probabilities, because disk masses come from a less strongly
varying section of the overall distribution. For example, $\sigma_{ln}
= 1$ yields 4\% and 8\% for 0.4\,$M_\odot$ and 1.5\,$M_\odot$ stars
respectively.

Our result is robust to changes in our assumed model parameters (in
the range 0.4--1.5\,$M_\odot$), because the range of core forming disk
masses generally remains the same, and the 6\% normalisation removes
absolute differences for different model parameters. 

Despite its simplicity, our model modified to have a constant snow
line at $a_{snow} = 2.7 \left( M_\star / M_\odot \right)^2$\,AU
produces a similar result to \citet{2005ApJ...626.1045I}: that less
gas giants form above a Solar mass. The difference arises from the
stronger dependence of snow line distance with stellar mass. For stars
$\lesssim$3\,$M_\odot$ in our model, accretion largely determines the
snow line distance, which suggests a better scaling is $a_{snow}
\propto M_\star^{4/9} \sigma_0^{2/9}$ if $\dot{M} \propto \sigma_0$
and $\sigma_g \propto M_\star \, a^{-3/2}$. Alternatively, $a_{snow}
\propto M_\star^{6/9}$ if $\dot{M} \propto M_\star$, or $a_{snow}
\propto M_\star^{8/9}$ if $\dot{M} \propto M_\star^2$.

Finally, our probability calculation does not take the increasing
core-forming region width with increasing stellar mass into
account. Cores that form later at larger distances may be less
susceptible to migrating into the central star, and stars with
multiple planets may be more likely to retain at least one during
migration and scattering processes. These effects have the potential
to increase the frequency of giant planets as stellar mass increases.

\section{Summary and Conclusions}\label{sec:summary}

We describe a model for the evolution of the snow line in a planet
forming disk, and apply it over a range of stellar masses to derive
the probability distribution of gas giants as a function of stellar
mass. The two main ingredients for our model are a prescription for
movement of the snow line due to accretion and PMS evolution, and
rules that determine whether protoplanets are massive enough, and form
early enough, to become gas giants.

The snow line distance generally moves inward over time. With our
prescription for the accretion rate, accretion dominates over
irradiation for stars with $M_\star \lesssim 2\,M_\odot$. For
$\gtrsim$3\,$M_\odot$ stars, irradiation dominates at times
$\gtrsim$1\,Myr as the star moves up to its main-sequence
luminosity. The transition is at a few Myr for $\sim$2\,$M_\odot$
stars. Over the wide range of observed accretion rates for any fixed
stellar mass, the snow line in some disks may be set entirely by
irradiation.

The snow line generally sets where the innermost gas giant cores
form. In relatively massive disks around intermediate mass stars,
rocky cores form interior to the snow line. The location of the
outermost core is always set by the gas dissipation timescale. The
range of disk masses that form cores, and the radial width of the
region in the disk where they form, increase with stellar mass. Lower
mass disks produce failed icy cores, which are probably similar to
Uranus, Neptune, and the observed ``super-Earths.''

The outward movement of the snow line as stars more massive than the
Sun reach the main-sequence, and as the disk becomes optically thin,
allows the ocean planets suggested by \citet{2004Icar..169..499L} to
form \emph{in situ}. The change in disk temperature is only large
enough for these planets to harbour oceans around stars
$\gtrsim$2.5\,$M_\odot$.

Our model includes several poorly determined parameters, which current
and future facilities will investigate. While there are current
resolved studies of gaseous disks \citep[e.g.][]{2007arXiv0704.1481B},
the next generation of telescopes such as GMT and ALMA will provide
more information on surface density profiles, and how disk properties
change with stellar mass and age. These studies will help to constrain
input parameters for our model. We have shown that the time dependence
of the snow line in part determines where gas giant cores form. This
result should motivate future studies of planet formation in disks
whose properties change with time.

The subsequent evolution of isolated cores is beyond the scope of this
paper, but further work that investigates the growth and dynamical
evolution of these objects can investigate the diversity of resulting
system structures.

Given an initial distribution of disk masses, the probability that a
star has at least one gas giant increases linearly with stellar mass
from 0.4\,$M_\odot$ to 3\,$M_\odot$. If the frequency of gas giants
around solar-mass stars is 6\%, we predict an occurrence rate of
1\% (10\%) for 0.4\,$M_\odot$ (1.5\,$M_\odot$)
stars. This result is largely insensitive to changes in our model
parameters.

In contrast to the \citet{2005ApJ...626.1045I} model, where it is hard
to form observable gas giants above 1\,$M_\odot$, our model predicts a
peak at $\sim$3\,$M_\odot$ because we include disk and PMS evolution
in our snow line derivation. However, our model does not include
migration, so our prediction applies to observable and currently
undetectable gas giants.

Though sample numbers are small, it appears that observable gas giant
frequency increases with stellar mass across a wide range of host
masses \citep{2007arXiv0707.2409J}. Larger samples of stars that host
giant planets, particularly low and intermediate-mass stars, will
solidify this result. These studies, and the extension of the results
to a wider range of semi-major axes, will provide a basis for
comparison with our model predictions.

\acknowledgements

We acknowledge support from an Australian Postgraduate Award, a
Smithsonian Astrophysical Observatory pre-doctoral fellowship (GK),
and the {\it NASA Astrophysics Theory Program} through grants
NAG5-13278 and NNG06GH25G (SK). We thank T. Currie, J. Johnson and the
ANU Planetary Science Institute planet group for helpful
discussions. We thank the anonymous referee for comments that improved
the manuscript.

\end{document}